\documentclass[%
 reprint,
 amsmath,amssymb,
 aps,
floatfix,
]{revtex4-1}

\usepackage{comment}

\usepackage[dvipdfmx]{graphicx}
\usepackage{dcolumn}
\usepackage{bm}

\begin{document}

\preprint{APS/123-QED}

\title{Many-variable variational Monte-Carlo study of superconductivity \\in two-band Hubbard models with an incipient band}

\author{Daichi Kato}
\author{Kazuhiko Kuroki}%
 \email{kuroki@phys.sci.osaka-u.ac.jp}
\affiliation{%
Department of Physics, Osaka University,\\
1-1 Machikaneyama, Toyonaka, Osaka 560-0043, Japan
}%

\date{\today}

\begin{abstract}
We study superconductivity in two-band models where one of the bands does or does not intersect the Fermi level depending on the parameter values. 
Applying a many-variable variational Monte-Carlo method for a Hubbard model on two-leg ladder and bilayer square lattices, 
we show that superconductivity can be enhanced in a parameter regime where the edge of one of the bands is near the Fermi energy, that is, when the band is incipient. 
The resemblence of the present results to those obtained by a weak coupling method in a recent study suggests that, even in the large $U$ regime, the suppression of the near-zero-energy spin fluctuations and the development of finite energy spin fluctuations are the key factors for the enhancement of superconductivity by an incipient band. 
\end{abstract}

\maketitle



\section{INTRODUCTION}
Purely electronic mechanism of superconductivity is expected to exhibit extremely high $T_\textrm{c}$ due to the large energy scale of the pairing glue originating from quantum fluctuations, such as spin and charge, orbital fluctuations.
Especially, spin-fluctuation-mediated pairing is one of the leading candidate mechanisms at work in unconventional high temperature superconductors, namely, cuprates and iron-based superconductors. 

In particular, in the early days of the study of the iron-based superconductors, it was considered that the Fermi surface nesting between electron and hole Fermi surfaces, combined with Hubbard $U$, induces spin fluctuations, which in turn act as pairing interaction around a certain wave vector $\bm{Q}$ if the gap sign changes across $\bm{Q}$. This kind of superconducting gap is referred to as the $s^\pm$ pairing~\cite{Hirschfeld_2011,KK_2013,Chubukov_2015,Hosono_2015,Mazin_2008,KK_2008}. 
However, the spin-fluctuation theory has been challenged by the discovery of (heavily) electron-doped iron-based superconductors with a relatively high $T_\textrm{c}$ where hole-like bands sink below the Fermi level leaving only electron-like Fermi surfaces~\cite{Guo_2010,Qian_2011,Wang_2012,Tan_2013,Liu_2012,He_2013,Lee_2014,Iimura_2012,Miao_2015,Niu_2015,Charnukha_2015,Miyata_2015}. Naively, removing the hole pocket is expected to destroy the spin-fluctuation-mediated pairing interaction and suppress $T_\textrm{c}$ rapidly. 

After these observations, ``incipient bands'', which sit close to, but do not intersect the Fermi level, have received much attention. Various authors have suggested that the spin-fluctuation scattering of pairs between an electron Fermi surface and an incipient hole band can induce $s^\pm$ pairing~\cite{Hirschfeld_2011,Wang_2011,bang_2014,Miao_2015,Charnukha_2015,Chen_2015,Linscheid_2016,Bang_2016,Bang_2019}. In this context, the bilayer Hubbard model, which has been extensively studied in the past~\cite{Bulut_1992,Scalettar_1994,Hetzel_1994,Santos_1995,Liechtenstein_1995,KK_2002,Kancharla_2007,Bouadim_2008,Lunata_2009,Zhai_2009} has recently attracted renewed focus. Having hole and electron Fermi surfaces, it can be regarded as a single-orbital analogue of the iron-based superconductors. In fact, it has been found in previous studies that $s^\pm$ pairing is favored over $d_{x^2-y^2}$ pairing by increasing the relative strength of the inter-layer nearest hopping to the intra-layer nearest hopping\cite{Maier_2011}. Further, as one of the bands becomes shallow or incipient, the spectral weight of spin-fluctuation is transferred to higher energies, which can lead to $s^\pm$ pairing state in which a gap appears on the hole band with the opposite sign to the gap on the electron Fermi surfaces~\cite{Mishra_2016,Nakata_2017,Maier_2019}. 

Regarding the incipient band situation, it was proposed in ref.~\cite{KK_2005} by one of the present authors and his coworkers that strongly enhanced superconductivity can take place in a system with coexisting wide and narrow bands when the narrow band sits in the vicinity of the Fermi level. There, the Hubbard model on a two-leg ladder was studied within the fluctuation exchange (FLEX) approximation. In the two-leg ladder model, which is a two-band model with bonding and antibonding bands, one of the bands becomes wide and the other becomes narrow when diagonal hoppings are introduced. In nowadays' terminology, the narrow band in this case is incipient. Quite recently, partially motivated by studies on various lattice models with coexisting wide and narrow (or flat) bands\cite{Kobayashi_2016,Misumi_2017,Ogura_2017,Matsumoto_2018,Sayyad_2019,Aoki_2019},  one of the present authors and his coworkers studied the bilayer model with diagonal interlayer hoppings\cite{Matsumoto_2019}, where one of the bands becomes wide and the other narrow, as in the two-leg ladder. There again, it has been shown using the FLEX approximation  that $s^\pm$-wave superconductivity is strongly enhanced when one of the bands is incipient. The role of the finite and low energy spin fluctuations on superconductivity, along with the commonalities and differences with the two-leg ladder, has been discussed~\cite{Matsumoto_2019}.

The above mentioned studies on the two-leg ladder and bilayer lattices with diagonal hoppings adopted the FLEX approximation\cite{KK_2005,Matsumoto_2018,Matsumoto_2019}, but because FLEX is basically a weak-coupling method, it is not clear whether the method can be applied to regimes where the electron-electron interaction is large. 
In the present study, we study Hubbard models on the two-leg ladder and bilayer square lattices, using a many-variable variational Monte-Carlo (mVMC) method~\cite{Tahara_2008,Misawa_2019}, which can be considered as reliable in the strong coupling regime\cite{Kainth_2019}. By comparing the results for the two-leg ladder (one dimensional) and the bilayer lattice (two dimensional), and with and without the diagonal hoppings, we discuss how the density of states (DOS) affects superconductivity and antiferromagnetism when one of the bands is close to being incipient.

\begin{figure}[tb]
\centering
\includegraphics[width=8cm]{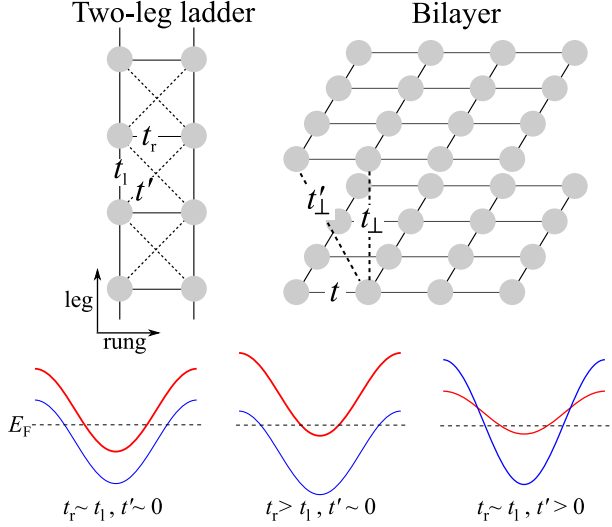}
\caption{Upper panel: the two-leg ladder lattice (left) and the bilayer lattice (right). Lower panels: typical band structures of the two-leg ladder lattice. Left: $t_\textrm{r}\sim t_\textrm{l}, t'\sim 0$, middle: $t_\textrm{r}> t_\textrm{l}, t'\sim 0$, right: $t_\textrm{r}\sim t_\textrm{l}, t'>0$.}
\label{models}
\end{figure}

\section{MODELS, METHOD, AND DEFINITIONS OF PHYSICAL QUANTITIES}
We study Hubbard models on the two-leg ladder and bilayer square lattices (Fig. \ref{models}). The Hamiltonian for the two-leg Hubbard ladder is
\begin{align}
H&=-t_\textrm{l}\sum_{\langle ij\rangle m\sigma}(c^{\dagger}_{im\sigma} c_{jm\sigma}+\textrm{h.c.})
-t_\textrm{r}\sum_{i\sigma}(c^{\dagger}_{i0\sigma} c_{i1\sigma}+\textrm{h.c.})\notag\\
&-t'\sum_{\langle ij\rangle \sigma}(c^{\dagger}_{i0\sigma} c_{j1\sigma}+\textrm{h.c.})
+U\sum_{im}n_{im\uparrow}n_{im\downarrow}.
\end{align}
Here $c_{im\sigma}^\dagger/c_{im\sigma}$ creates/annihilates a fermion with spin $\sigma(=\uparrow,\downarrow)$ on the $i^\textrm{th}$ site  on the $m^\textrm{th}$ chain ($m$=0 or 1) and $n_{im\sigma}=c_{im\sigma}^\dagger c_{im\sigma}$. The nearest neighbor hoppings in the leg and rung directions are $t_\textrm{l}$ and $t_\textrm{r}$, respectively, and the next nearest neighbor diagonal hopping is $t'$. Since $t_\textrm{r}$ connects two chains, we will call this the inter-chain hopping. The band structure for this model is
\begin{align}
\varepsilon(\bm{k})&=-2(t_\textrm{l}+ t'\cos k_y)\cos k_x-t_\textrm{r}\cos k_y,
\end{align}
where the case of $k_y=0\ (\pi)$ corresponds to the bonding (anti-bonding) band. 
For $t'>0$, the bonding band is wide and the anti-bonding band is narrow.

The Hamiltonian for the square lattice bilayer Hubbard model is
\begin{align}
H&=-t\sum_{\langle ij\rangle m\sigma}(c^{\dagger}_{im\sigma} c_{jm\sigma}+\textrm{h.c.})
-t_{\perp}\sum_{i\sigma}(c^{\dagger}_{i0\sigma} c_{i1\sigma}+\textrm{h.c.})\notag\\
&-t'_{\perp}\sum_{\langle ij\rangle \sigma}(c^{\dagger}_{i0\sigma} c_{j1\sigma}+\textrm{h.c.})
+U\sum_{im}n_{im\uparrow}n_{im\downarrow}.
\end{align}
Here $c_{im\sigma}^\dagger/c_{im\sigma}$ creates/annihilates a fermion with spin $\sigma(=\uparrow,\downarrow)$ on the $i^\textrm{th}$ site  on the $m^\textrm{th}$ layer ($m$=0 or 1).
The intra-layer hopping is $t$ and the inter-layer hopping is $t_\perp$, the next nearest neighbor inter-layer hopping is $t'_\perp$. The band structure for this model is
\begin{align}
\varepsilon(\bm{k})&=-2(t+t'_\perp\cos k_z)(\cos k_x+\cos k_y)-t_\perp\cos k_z,
\end{align}
where the case of $k_z=0\ (\pi)$ corresponds to the bonding (anti-bonding) band. 
For $t'_\perp>0$, the bonding band is wide and the anti-bonding band is narrow.

We take $N_\textrm{s}=60\times 2\ (12\times 12\times 2)$ sites for the two-leg ladder (bilayer) Hubbard model with the antiperiodic-periodic boundary condition in $x\ (y)$ direction. The band filling is defined as $n=N_\textrm{e}/N_\textrm{s}$ where $N_\textrm{e}=\sum_{mi\sigma}n_{im\sigma}$. Hereinafter, the site index $(i,m)$ is simply written as $i$.

To study the ground state of these Hubbard models, we employ a mVMC method~\cite{Tahara_2008,Misawa_2019}, which can describe the strong correlation and various ordering fluctuations accurately. Our variational wave function is defined as
\begin{equation}
|\phi\rangle={\cal P}_\textrm{G}{\cal P}_\textrm{J}|\phi_\textrm{pair}\rangle,
\end{equation}
where ${\cal P}_\textrm{G}, {\cal P}_\textrm{J}$ are the Gutzwiller and Jastrow correlation factors, respectively. The Gutzwiller factor punishes the double occupation of electrons defined as
\begin{equation}
{\cal P}_\textrm{G} = \exp\left(-\frac{1}{2}\sum_ig_in_{i\uparrow}n_{i\downarrow}\right).
\end{equation}
The Jastrow factor is defined as 
\begin{equation}
{\cal P}_\textrm{J} = \exp\left(-\frac{1}{2}\sum_{ij}v_{ij}n_in_j\right),
\end{equation} 
where $n_i=\sum_{\sigma}n_{i\sigma}$. The long-range part of this factor drives the distinction between the metal and insulator. $|\phi_\textrm{pair}\rangle$ is the one-body part defined as
\begin{equation}
|\phi_\textrm{pair}\rangle = \left[\sum_{i, j=1}^{N_\textrm{s}}f_{ij}c_{i\uparrow}^{\dag}c_{j\downarrow}^{\dag} \right]^{N_\textrm{e}/2}|0 \rangle \label{fij},
\end{equation}
where $f_{ij}$ is assumed to have $2\times2\ (2\times2\times2)$ sublattice structure or equivalently $2\times2\times N_\textrm{s}\ (2\times2\times2\times N_\textrm{s})$ independent variational parameters for one-body part in the two-leg ladder (bilayer) systems. 

To study a possible superconducting state, we consider the following BCS wave function,
\begin{align}
|\phi_\textrm{BCS} \rangle &= \left(\sum_{\bm{k}\in \textrm{BZ}}\varphi(\bm{k})c^{\dag}_{\bm{k}\uparrow}c^{\dag}_{\bm{-k}\downarrow}\right)^{N_\textrm{e}/2}|0\rangle,
\end{align}
with
\begin{align}
\varphi(\bm{k})&=\frac{\Delta(\bm{k})}{\xi(\bm{k})+\sqrt{\xi(\bm{k})^2+\Delta(\bm{k})^2}},
\end{align}
where $\xi(\bm{k})=\varepsilon(\bm{k})-\mu$, $\mu$ is the chemical potential and $\Delta(\bm{k})$ is the superconducting gap. The BCS wave function is rewritten in the real space representation as follows: 
\begin{equation}
f_{ij}=\frac{1}{N_\textrm{s}}\sum_{\bm{k}}\varphi(\bm{k})\exp\left[i\bm{k}\cdot(\bm{r}_i-\bm{r}_j)\right].
\end{equation}

In this study, we employ the BCS partial $d\ (s^\pm)$-wave superconducting state as the initial states for the ladder (bilayer) system, namely, $\Delta(\bm{k})=\Delta_0\cos k_y\ (\Delta_0\cos k_z)$. The variational parameters are simultaneously optimized to minimize the variational energy by using the stochastic reconfiguration method~\cite{Sorella_2001}.

To investigate the ground state properties of these Hubbard models, we calculate the momentum distribution function and spin-structure factor, equal-time superconducting correlation. The momentum distribution function is defined as
\begin{align}
n_\sigma(\bm{q})&=\frac{1}{N_\textrm{s}}\sum_{i,j}\langle c_{i\sigma}^\dagger c_{j\sigma}\rangle \exp \left[i\bm{q}\cdot\left(\bm{r}_i-\bm{r}_j\right)\right].
\notag
\end{align}
and the spin-structure factor is defined as
\begin{align}
S(\bm{q})&=\frac{1}{3N_\textrm{s}}\sum_{i,j}\langle \bm{S}_i\cdot\bm{S}_j\rangle \exp \left[i\bm{q}\cdot\left(\bm{r}_i-\bm{r}_j\right)\right].
\notag
\end{align}
Further, the equal-time superconducting correlations are defined as
\begin{equation}
P_\alpha(\bm{r})=
\frac{1}{2N_\textrm{s}}
\sum_{\bm{r}_i}
\langle
\Delta^{\dagger}_{\alpha}(\bm{r}_i)
\Delta_{\alpha}(\bm{r}_i+\bm{r})
+
\Delta_{\alpha}(\bm{r}_i)
\Delta^{\dagger}_{\alpha}(\bm{r}_i+\bm{r})
\rangle.
\notag
\label{SCC}
\end{equation}
Superconducting order parameters $\Delta_{\alpha}(\bm{r}_{i})$ are defined as
\begin{equation}
\Delta_{\alpha}(\bm{r}_{i})=\frac{1}{\sqrt{2}}\sum_{\bm{r}}
f_{\alpha}(\bm{r})
(
c_{\bm{r}_i\uparrow}
c_{\bm{r}_i+\bm{r}\downarrow}
-
c_{\bm{r}_i\downarrow}
c_{\bm{r}_i+\bm{r}\uparrow}
).
\notag
\end{equation}
Here $f_{\alpha}(\bm{r})$ is the form factor that describes the symmetry of the superconductivity. For the partial d-wave superconductivity in two-leg ladder systems, we define
\begin{equation}
f_d(r_x,r_y)=\delta_{r_x,0}\delta_{r_y,1},
\notag
\end{equation}
where $\delta_{ij}$ denotes the Kronecker's delta. 
For the $s^\pm$-wave superconductivity in bilayer systems, we define
\begin{equation}
f_{s^\pm}(r_x,r_y,r_z)=\delta_{r_x,0}\delta_{r_y,0}\delta_{r_z,1}.
\notag
\end{equation}

To reduce stochastic errors, we calculate long-range average of the superconducting correlation, which is defined as
\begin{equation}
\overline{P}_{\alpha}=
\frac{1}{M}\sum_{2<|\bm{r}|<r_\textrm{max}}
P_{\alpha}(\bm{r}),
\notag
\end{equation}
where $r_\textrm{max}$ is $30\ (6\sqrt{2})$ for the present two-leg ladder (bilayer) models.
$M$ is the number of vectors satisfying $2<r<r_\textrm{max}$. 
Here, we eliminate the short range part of the superconducting correlation since it does not reflect the off-diagonal ordering nature of superconductivity 
in order to reduce the effect of the boundary condition.

\section{RESULTS}
\subsection{Two-leg Hubbard ladder}

\begin{figure*}[tb]
\centering
\includegraphics[width=14cm]{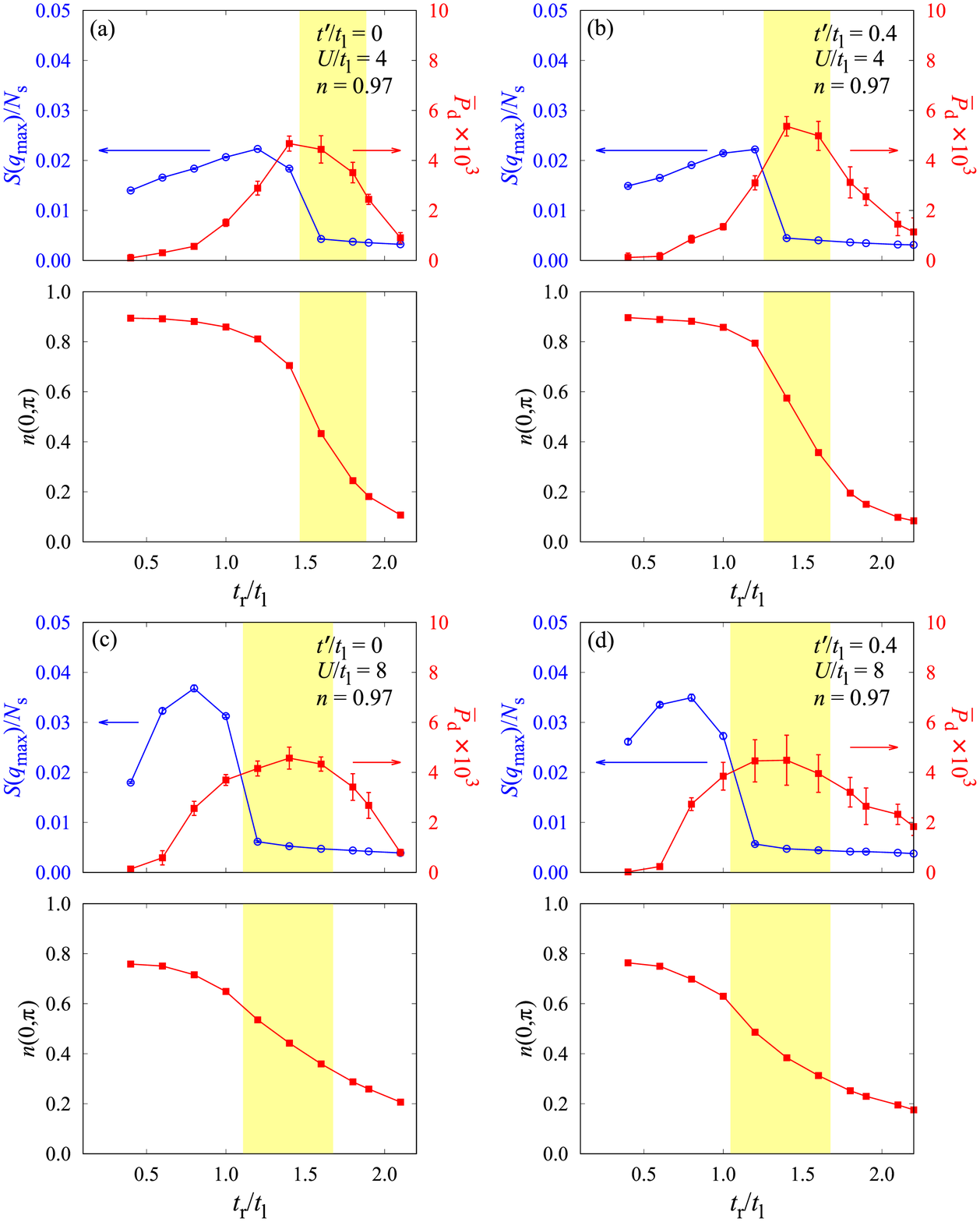}
\caption{(color online). Inter-chain hopping $t_\textrm{r}/t_\textrm{l}$ dependence of the averaged partial $d$-wave superconducting correlation $\overline{P}_d$ and the peak value of the spin structure factor $S(\bm{q}_{\textrm{max}})$ (upper panels), the momentum distribution function of the anti-bonding band minimum $n(0,\pi)$ (lower panels) for the two-leg ladder model with (a) $t'/t_\textrm{l}=0$ and $U/t_\textrm{l}=4$, (b) $t'/t_\textrm{l}=0.4$ and $U/t_\textrm{l}=4$, (c) $t'/t_\textrm{l}=0$ and $U/t_\textrm{l}=8$, (d) $t'/t_\textrm{l}=0.4$ and $U/t_\textrm{l}=8$. The band filling is $n=0.97$. The yellow region denotes the incipient-band regime. In the present plots and the plots in the later figures, the error bars indicate the estimated statistical errors of the Monte Carlo sampling.}
\label{ladder}
\end{figure*}

We begin with the two-leg ladder. Figure \ref{ladder}a shows the inter-chain hopping dependence of several physical properties for $t'/t_\textrm{l}=0$ and $U/t_\textrm{l}=4$; peak value of the spin structure factor $S(\bm{q}_{\textrm{max}})$, which is the square of the anti-ferromagnetic ordered moment, and average value of superconducting correlation $\overline{P}_d$ at long distance with the partial $d$ symmetry, corresponding to the square of the superconducting order parameter. We also plot the momentum distribution function at the anti-bonding band minimum $n(0,\pi)$, which monitors whether or not the anti-bonding band intersects the Fermi level. For $1.2\leq t_\textrm{r}/t_\textrm{l}\leq1.6$, $n(0,\pi)$ decreases rapidly and $S(\bm{q}_{\textrm{max}})$ is strongly suppressed as $t_\textrm{r}/t_\textrm{l}$ increases. Therefore, the incipient-band regime is estimated to be in a range of $1.2\leq t_\textrm{r}/t_\textrm{l}\leq1.6$.
In the incipient-band regime, $\overline{P}_d$ is maximized.
Further, the partial $d$-wave superconducting phase exhibits the dome structure around $t_\textrm{r}/t_\textrm{l}\sim 1.5$, which is consistent with a previous FLEX study on the two-leg ladder Hubbard model without $t'$~\cite{KK_2001}. 
For $t'/t_\textrm{l}=0.4$, where there are wide and narrow bands, the inter-chain hopping dependence of several physical properties are similar to those for $t'/t_\textrm{l}=0$ as shown in Fig. \ref{ladder}b. 
For a larger interaction value of $U/t_\textrm{l}=8$, the variation of $n(0,\pi)$ against $t_\textrm{r}/t_\textrm{l}$ becomes broad due to correlation effects as shown in Figs. \ref{ladder}c and \ref{ladder}d.
On the other hand, we find a clear suppression of $S(\bm{q}_{\textrm{max}})$, which indicates the Lifshitz transition. 
Thus, superconductivity is optimized when a band becomes incipient also in the strongly correlated regime.

\begin{figure*}[tb]
\centering
\includegraphics[width=14cm]{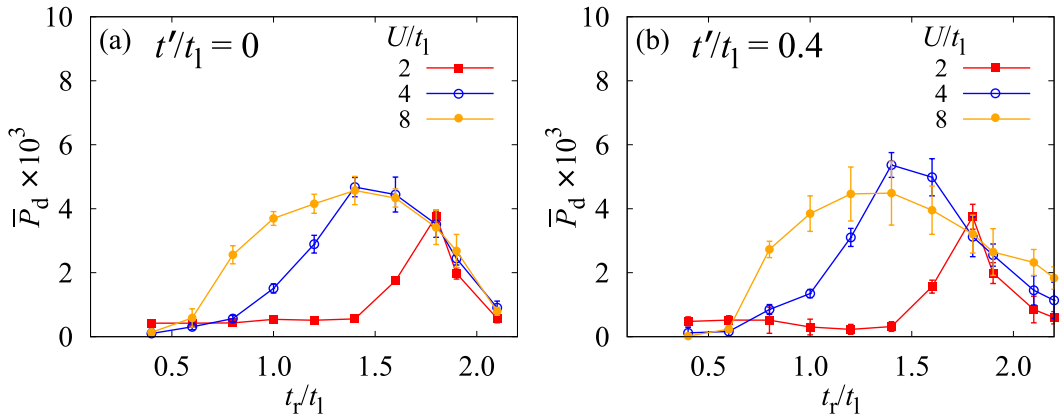}
\caption{(color online). $t_\textrm{r}/t_\textrm{l}$ dependence of $\overline{P}_d$ for the two-leg ladder model with various values of $U/t_\textrm{l}$ and (a) $t'/t_\textrm{l} = 0$, (b) $t'/t_\textrm{l} = 0.4$. The band filling is $n=0.97$. }
\label{ladder2}
\end{figure*}

We also study $t_\textrm{r}/t_\textrm{l}$ dependence of the superconducting correlation $\overline{P}_d$ for various values of $U/t_\textrm{l}$ as shown in Fig. \ref{ladder2}. 
Both for $t'/t_\textrm{l}=0$ and $t'/t_\textrm{l}=0.4$, the regime of enhanced superconducting correlation extends to smaller $t_\textrm{r}/t_\textrm{l}$ as $U/t_\textrm{l}$ increases, presumably due to band narrowing caused by $U$.
In general, Hubbard $U$ can narrow bands near the Fermi level and induce the Lifshitz transition\cite{Ogura_2019}. 

\begin{figure*}[tb]
\centering
\includegraphics[width=14cm]{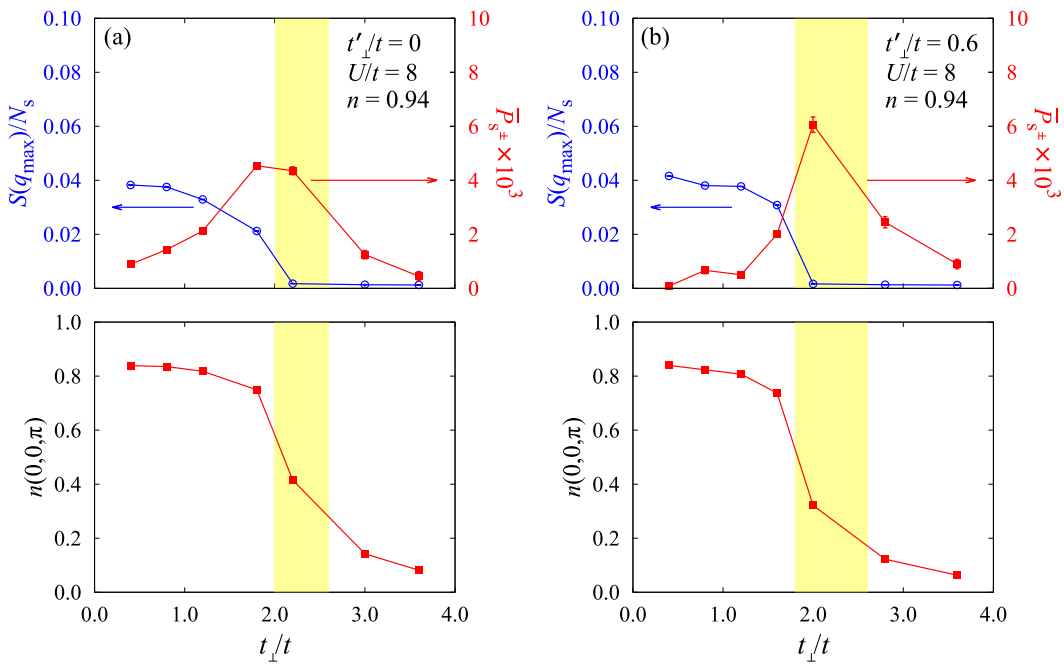}
\caption{(color online). Inter-layer hopping $t_\perp/t$ dependence of the averaged $s^\pm$-wave superconducting correlation $\overline{P}_{s^\pm}$ and the peak value of the spin structure factor $S(\bm{q}_{\textrm{max}})$ (upper panels), the momentum distribution function of the anti-bonding minimum $n(0,0,\pi)$ (lower panels) for the bilayer model with (a) $t'_\perp/t=0$ and $U/t=8$, (b) $t'_\perp/t=0.6$ and $U/t=8$. The band filing is $n=0.94$.}
\label{bilayer1}
\end{figure*}

\begin{figure*}[tb]
\centering
\includegraphics[width=14cm]{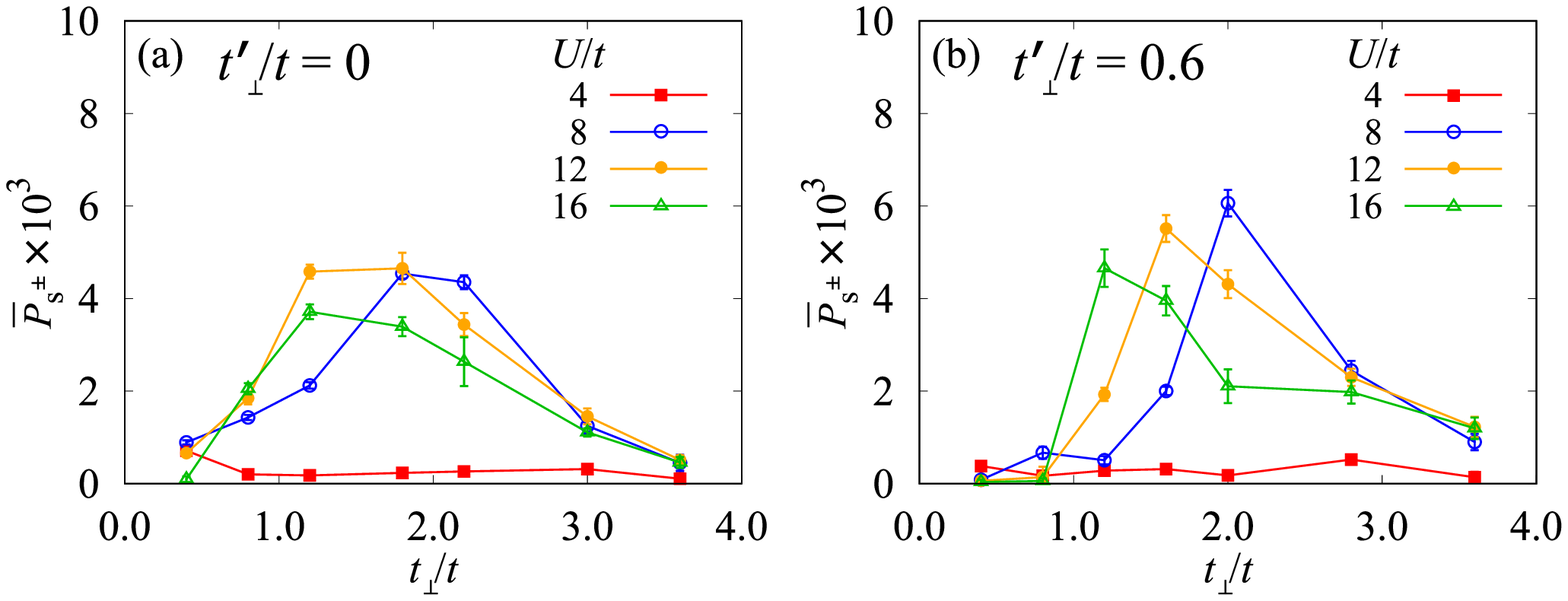}
\caption{(color online). $t_\perp/t$ dependence of $\overline{P}_{s^\pm}$ for the bilayer model with various values of $U/t$ and (a) $t'_\perp/t = 0$, (b) $t'_\perp/t = 0.6$. The band filling is $n=0.94$. }
\label{bilayer2}
\end{figure*}

\subsection{Bilayer Hubbard model}
We next move on to the bilayer model. Figure \ref{bilayer1}a shows the inter-layer hopping dependence of  physical properties for $t'_\perp/t=0$ and $U/t=8$; the peak value of the spin structure factor $S(\bm{q}_{\textrm{max}})$ and the average value of the superconducting correlation $\overline{P}_{s^\pm}$ at long distances with the s$^\pm$ symmetry. We also plot the momentum distribution function at the anti-bonding band minimum $n(0,0,\pi)$.
For $t_\perp/t>1.8$, $n(0,0,\pi)$ decreases steeply, and $S(\bm{q}_{\textrm{max}})$ is strongly suppressed as $t_\perp/t$ increases.
Thus, the incipient-band regime is estimated to be in a range of $1.8\leq t_\perp/t \leq 2.4$. 
Around the incipient-band regime, $\overline{P}_{s^\pm}$ is enhanced.
The $s^\pm$-wave superconducting correlation exhibits a dome structure around $t_\perp/t\sim 2.0$~\cite{dwave}, which is consistent with FLEX~\cite{KK_2002} and fRG~\cite{Zhai_2009}, DCA~\cite{Maier_2011} studies. 
For $t'_\perp/t=0.6$, $t_\perp/t$ dependence of the physical properties are basically similar to those for $t'_\perp/t=0$ as shown in Fig. \ref{bilayer1}b. 

As in the two-leg ladder, we also study $t_\perp/t$ dependence of superconducting correlation $\overline{P}_{s^\pm}$ for various values of $U/t$ as shown in Fig. \ref{bilayer2}. 
Both for $t'_\perp/t=0$ and $t'_\perp/t=0.6$, the regime where the superconducting correlation develops extends to lower $t_\perp/t$ as $U/t$ increases in the same way as the two-leg Hubbard ladder. 

We note that $n(0,0,\pi)$ of the $U=8t$ bilayer Hubbard model varies steeper than $n(0,\pi)$ of the $U=8t_\textrm{l}$ two-leg Hubbard ladder, and actually resembles that of $U=4t_\textrm{l}$ ladder. Since the broadness of the momentum distribution variation around the Lifshitz transition presumably originates from the correlation effect, the present result indicates the strength of the electron correlation is roughly determined by $U/W$, where $W$ is the band width.


\section{DISCUSSION}
\begin{figure}[tb]
\centering
\includegraphics[width=7cm]{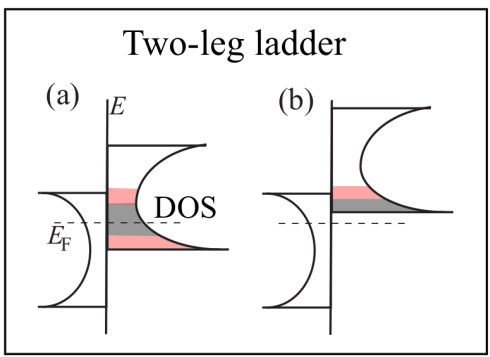}
\caption{Typical density of states of the two-leg ladder Hubbard model. 
In each figure, the left (right) side of the vertical line depicts
the DOS of the bonding (antibonding) band. The gray area denotes the
portion of the bonding band DOS which gives rise to the low-lying pair breaking
spin fluctuations, and the red area is the portion of the antibonding
band DOS contributing to the spin fluctuations which mediate pairings.
(a) when both bands intersect the Fermi level, and (b) when the antibonding band is incipient. 
}
\label{DOS_ladder}
\end{figure}

\begin{figure}[tb]
\centering
\includegraphics[width=7cm]{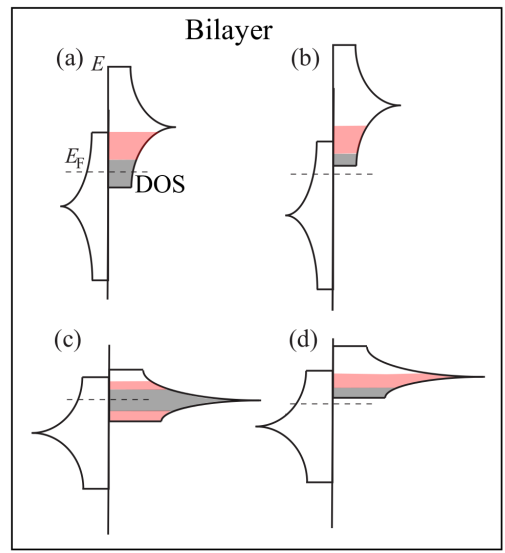}
\caption{Typical density of states for the bilayer Hubbard model. 
(a)(b) when $t'_\perp/t$ is (nearly) zero, and (c)(d) when $t'_\perp/t$ is finite. (a)(c) when both bands intersect the Fermi level, and (b)(d) when the antibonding band is incipient. 
}
\label{DOS_bilayer}
\end{figure}

The present results show that superconductivity is enhanced in the incipient band regime regardless of whether the system is one- or two-dimensional, or whether one of the bare bands is narrow or not. This is in fact consistent with the recent FLEX studies\cite{Matsumoto_2018,Matsumoto_2019,MatsumotoComment}.
In this section, we further discuss the relation between the bilayer and two-leg ladder models, based on observations made in previous studies.
Refs.~\cite{Mishra_2016,Nakata_2017,Matsumoto_2019} pointed out the important role of the finite-energy spin-fluctuations played in the enhancement of superconductivity in the bilayer Hubbard model. 
Especially, quite recently, in ref.\cite{Matsumoto_2019},  the role played by the spin fluctuations  in different energy ranges in two-band systems has been discussed as follows.
There is a critical frequency $\omega_\textrm{c}$ in the spin-fluctuation spectrum, which should be smaller than a pairing cutoff energy $\varepsilon_\textrm{c}$. 
In general, the low-energy spin-fluctuation with $\omega<\omega_\textrm{c}$ leads to strong renormalization and hence are ``pair breaking" whilst the finite-energy spin fluctuations with $\omega>\omega_\textrm{c}$ enhance $T_\textrm{c}$~\cite{Millis_1988}. Thus, when the low-lying spin fluctuations are suppressed while the finite energy spin fluctuations are enhanced, superconductivity can be enhanced.
In multi-band systems, as one of the bands moves away from the Fermi level, the spin-fluctuation spectral weight is transferred to higher energies. 
When the spin-fluctuation spectral weight is away from the critical frequency of spin-fluctuations, but is within the paring cutoff energy ($\varepsilon_\textrm{c}>\varepsilon>\omega_\textrm{c}$), the pairing interaction can be strong without strongly renormalizing the quasiparticles. On the other hand, when the spin fluctuations are concentrated at very low or too high energies, 
superconductivity is degraded. From this viewpoint, we further discuss the relation between the superconducting correlation obtained by mVMC and the shape of the DOS of the antibonding band. 

One difference between the bilayer and the two-leg ladder observed in the present study is the $U$ dependence of superconductivity. Namely, in  the two-leg ladder, the superconducting correlation is enhanced even for $U/t_\textrm{l}=2$ when the antibonding band is incipient, while such an enhancement is not obtained for the bilayer model for $U/t=4$. Note that here we compare two cases where $U$ normalized by the bare band width are the same. Actually, a similar result was obtained in the recent FLEX calculation~\cite{Matsumoto_2019}. There, it has been pointed out that in the bilayer model, the correlation effect reduces the width of the incipient band\cite{Ogura_2019}, which makes more spin fluctuation weight lie within the energy regime effective for pairing. In the two-leg ladder, such effect is not necessary when the antibonding band is incipient because in a one-dimensional system, DOS is diverging at the band edge (see Fig. \ref{DOS_ladder}).

Another point that we notice, if we look closely, is that in the two-leg ladder, the superconducting correlation, maximized around the incipient-band regime, is reduced as $t_\textrm{r}/t_\textrm{l}$ decreases and the antibonding band intersects the Fermi level, but does so rather mildly and smoothly especially for $U=8t_\textrm{l}$, whereas the reduction of the superconducting correlation in the bilayer model upon reducing $t_\perp$ occurs rapidly after the antibonding band intersects the Fermi level. If we compare in more detail the two cases for the bilayer model, the reduction of the superconducting correlation is more abrupt for the case with $t'_\perp=0.6$. These differences can again be understood from the shape of the DOS (see Figs. \ref{DOS_ladder} and \ref{DOS_bilayer}). Namely, in the two-leg ladder, where the DOS at the band edge is diverging, the DOS at the Fermi level decreases (rapidly, especially for large $U$ because the band width shrinks due to renormalization) as $t_\textrm{r}$ is reduced after the antibonding band intersects the Fermi level, so that the pair-breaking low-energy spin fluctuations become weaker. By contrast, in the bilayer model, where the DOS is diverging around the middle of the antibonding band, the DOS at the Fermi level increases upon reducing $t_\perp$ after the antibonding Fermi surface is formed, resulting in an increase of the pair-breaking spin fluctuations. The diverging DOS of the antibonding band approaches the Fermi level ``faster'' when $t'_\perp$ is finite, so that the superconducting correlation is rapidly suppressed for the case of $t'_\perp=0.6$ as $t_\perp$ is reduced. A similar analysis has been performed in ref.~~\cite{Matsumoto_2019}, not for the $t_\textrm{r}, t_\perp$ variation, but for the $t'(t'_\perp)$ variation of superconductivity.

\section{SUMMARY}
To summarize, we have studied superconductivity in the Hubbard model on the two-leg ladder and bilayer square lattices.
In both systems, superconductivity can be optimized in a region around the Lifshitz point, where one of the bands is (nearly) incipient. The parameter dependence of the superconducting correlation function is reminiscent of the FLEX results obtained in ref.\cite{Matsumoto_2019}, which supports the scenario that superconductivity is enhanced by an incipient band due to the suppression of the near-zero-energy spin fluctuations and enhanced finite energy spin fluctuations working as an effective pairing glue. We stress that the consistency between the two approaches is highly nontrivial because the two approaches are totally different; FLEX is based on a weak coupling perturbational theory, which takes into account the spin fluctuations (in momentum space) explicitly in the effective interaction, whereas the present mVMC method takes into account the electron correlation effect in a real-space-based manner, which is expected to be more appropriate in the strong coupling regime. Since it has been shown that incipient bands enhance superconductivity in other models\cite{Matsumoto_2018,Kobayashi_2016,Aoki_2019,Sayyad_2019,Misumi_2017,Ogura_2017,Ogura_2019},  it is an interesting future problem to study those models using the mVMC method.

\section*{Acknowledgments}
We appreciate Takahiro Misawa and Kota Ido for providing us guidance in the mVMC method. We also thank Masayuki Ochi, Karin Matsumoto, and Daisuke Ogura for fruitful discussions. The numerical calculations were performed at the  following institutions; the Supercomputer Center, Institute for Solid State Physics, University of Tokyo and Yukawa Institute Computer Facility, Kyoto University, the Cybermedia Center, Osaka University.



\bibliography{refs}

\begin{thebibliography}{55}
\bibitem [{dwa()}]{dwave}%
  \BibitemOpen
  \href@noop {} {}\bibinfo {note} {The $d_{x^2-y^2}$ pairing dominates over the
  $s^\pm$ pairing only for small $t_\perp (\leq 0.8t)$ if $d$-wave states are
  employed as initial states.}\BibitemShut {Stop}%
\end{thebibliography}

\end{document}